*Review Article*

Title: Breaking Boundaries: A Chronology with Future Directions of Women in Exercise Physiology Research, Centred on Pregnancy

**Abbey E. Corson[a], Meaghan MacDonald[a], Velislava Tzaneva[a], Chris M. Edwards[a], *Kristi B. Adamo[a,b]**

[a]School of Human Kinetics, Faculty of Health Sciences, University of Ottawa
[b]Faculty of Medicine, Obstetrics and Gynecology
* Corresponding author: Kristi B. Adamo (kadamo@uottawa.ca)

Present Address:   University of Ottawa – Lees Campus
200 Lees Avenue, Ottawa, ON
K1S 5L5

**Abstract:**
Historically, females were excluded from clinical research due to their reproductive roles, hindering medical understanding and healthcare quality. Despite guidelines promoting equal participation, females are underrepresented in exercise science, perpetuating misconceptions about female physiology. Even less attention has been given to exercise in the pregnant population. Research on pregnancy and exercise has evolved considerably from the initial bedrest prescriptions but concerns about exercise risks during pregnancy persisted for many decades. Recent guidelines endorse moderate-intensity physical activity during pregnancy, supported by considerable evidence of its safety and benefits. Mental health during pregnancy, often overlooked, is gaining traction, with exercise showing promise in reducing depression and anxiety. While pregnancy guidelines recommend moderate-intensity physical activity, there remains limited understanding of optimal frequency, intensity, type and time (duration) for extremes like elite athletes or those with complications. Female participation in elite sport and physically demanding jobs is rising, but research on their specific needs is lacking. Traditional practices like bed rest for high-risk pregnancies are being questioned, as evidence suggests it may not improve outcomes. Historical neglect of gestational parents in research perpetuated stereotypes of female frailty, but recent years have seen a shift towards recognizing the benefits of an active pregnancy. Closing knowledge gaps and inclusivity in research are crucial for ensuring guidelines reflect the diverse needs of gestational parents. Therefore, the purpose of this review is to summarize the evolution of exercise physiology and pregnancy research along with future directions for this novel field.

# 1. Introduction

Historically, females have been viewed as reproductive vessels and the existence of reproductive organs led to their exclusion from clinical decision-making research. This exclusion was concomitant with a pervasive status quo that places a lower value on females as a result of conventional, cultural and systemic forces that stigmatize their participation in scientific research[1,2]. Owing to hormonal and reproductive organ complexities, it is common for researchers to adopt a very cautious approach to including females in clinical trials and scientific research. Since important medical research did not include representation of females and people with menstrual cycles, the quality of healthcare accessible to them has suffered[3].

In recent decades, it has become apparent that excluding female participants from scientific research has hindered our understanding of their disease risk and progression as well as responsiveness to medications and other therapeutic treatments (e.g., medical devices and natural health products). Many of the breakthroughs in medicine stem from research primarily conducted on male cells and animals with the results of these studies being subsequently extrapolated to females. Consequently, the practice of applying knowledge gathered on males to females has impacted the pharma industry, with eight out of ten approved drugs being withdrawn from the market due to unforeseen health implications in female users[4]. Another example refers to the key differences between sexes surrounding cardiovascular disease (CVD) risk. Evidence of this disparity goes back to the late 1950's[5]. It is well documented now that females die from CVD more frequently than men, yet it remains understudied, under-recognized, underdiagnosed, and undertreated in the female population [6].

Not only have males predominantly filled the roles of doctors, scientists, and researchers, but they have been the focus of the majority of the medical science studies on significant health issues that affect all sexes, like hypertension and diabetes. While clear lifestyle, environmental and behavioural proclivities between males and females that are reflected in biological differences at the molecular and cellular level, it was not until 1994 that the US National Institutes of Health (NIH) created a guideline for the evaluation of sex-based differences in clinical trials to assess the safety and efficacy of drugs to treat the masses [7]. In Canada, concerns raised about a deficit of research on breast and gynecological cancers, led to the development of the "Canadian Guidance Document on The Inclusion of Women in Clinical Trials" recommending the inclusion of females at each stage of scientific research so that the impact of new drugs on specific sexes can be better understood[8].

The issue of female representation is not limited to medical or clinical research. Despite guidelines aimed at promoting equal participation by both sexes, inclusion of females in exercise sciences remains inadequate. Cowley *et al,* drew attention to the disproportions in their examination of the ratio between male and female participants in sport science research; 34% of the study population were female, and only 6% of total publications focused exclusively on females[9]. Furthermore, a recent systematic review examining the prevalence of female participants in exercise studies focusing on vascular endothelial function noted that only 36% of participants were female[10]. The scarce literature on how females respond to exercise, exercise training, and other physical activity (PA) interventions has led to a lack of understanding on this topic relative to men[11], as visualized in Figure 1. The factors leading to the low participation of females in exercise and physiology research are akin to those preventing them from participating in medical trials. Namely, concerns and uninformed beliefs about female physiology such as hormonal fluctuations during the menstrual cycle, hormonal contraceptives, and historical beliefs that females are not capable or interested in exercise. Whilst the common assumption that females are smaller versions of males continues to be the subject of debate in the peer-reviewed literature, with the publication of her book - *Up to Speed* - Christine Yu[12] is shedding light in the public sphere on the detrimental impact of applying exercise physiology and nutrition research results from males to females in athletics. The lack of evidence-informed practices with female athletes has led to the disruption of menses, stress fractures, drop-out and much more[13].

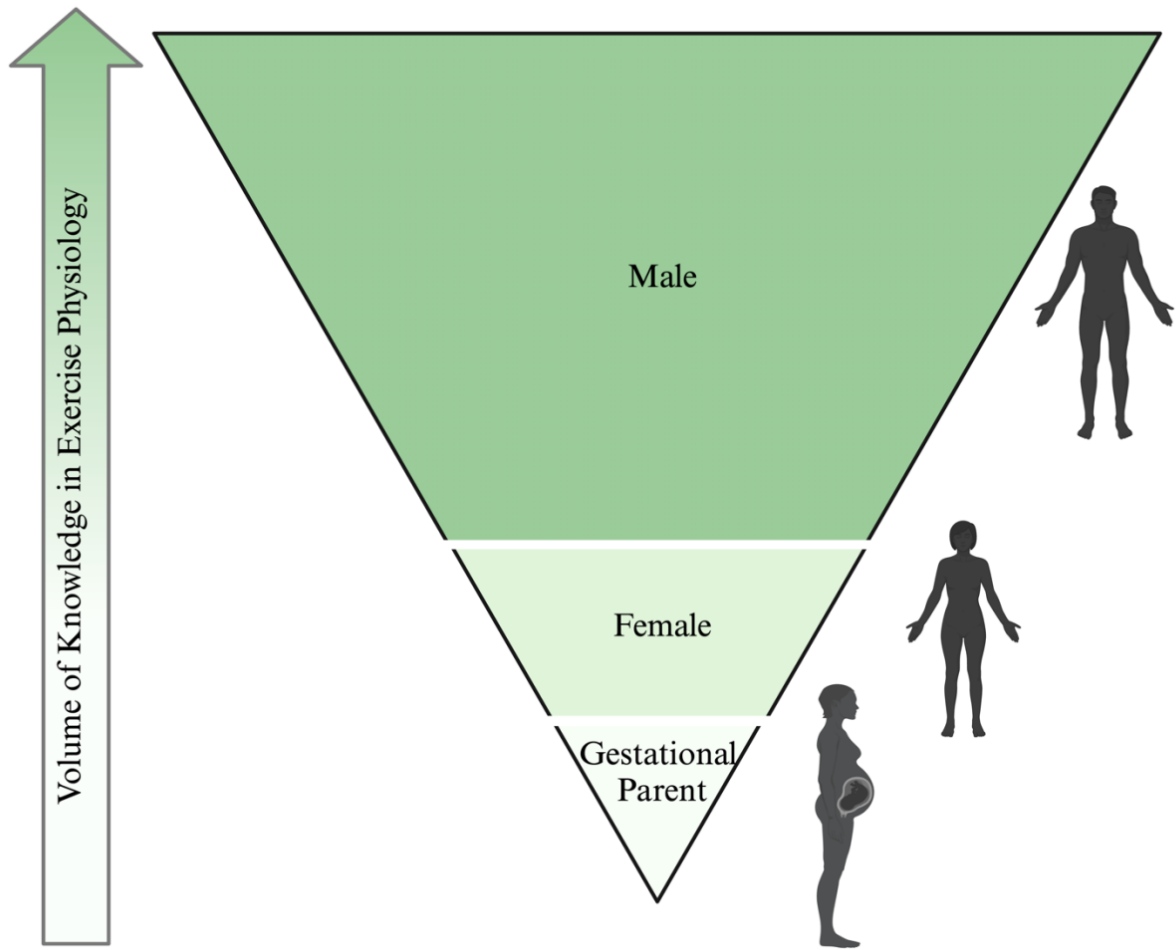

Figure 1. Visualization of the Approximate Volume of Knowledge in Exercise Physiology Research of the Biological Sexes and Gestational Parents. Created using BioRender.com

While females, in general, make up the minority of research subjects, an even greater research pariah, especially in exercise science, are pregnant females. Subsequently, our knowledge and understanding of physiological responses to exercise in this population is lagging. The fear of studying pregnancy is, in part, due to the therapeutic tragedies of the past (e.g., the use of Thalidomide in the 1960s and offspring limb deformities) [14] and the perception of vulnerability. Undoubtedly, certain cautions should be applied when engaging with the pregnant population, but the sizeable underrepresentation of gestational parents (gesPs) in exercise and sports sciences has been fueled by the fear that gesPs are not capable of engaging in exercise without impairing fetal growth and development. While today we know these assumptions are not factual, the inclusion of gesPs in exercise physiology research is still low. Moreover, publications regarding pregnancy and exercise physiology may marginalized due to the underrepresentation of females on sports science editorial boards, and fewer holding first and senior authorship positions in publications of randomized controlled trials compared to men[15]. Consequently, exercise physiology research focused on females may not be prioritized by some journals nor prioritized by male researchers.

As can be gathered from any exercise physiology or sports science textbook, research advancements in exercise science have been monumental in the last two centuries, with the male population and their responses being well characterized. While our understanding of females in exercise physiology research is gathering speed, the research remains significantly behind what we know about males, and our knowledge about the exercising pregnant population is still in the starting blocks. Therefore, this review aims to summarize the evolution of exercise physiology research on gesPs and offer potential future directions in this unique population.

## 2. Timeline
### 2.1. *1850s – Early 1900s: A Time for Bedrest*

For centuries, bedrest was a common pregnancy prescription touted by medical professionals. In 1858, a midwifery textbook, written by Fleetwood Churchill, described gestational rest as "the most powerful prophylactic means we possess" [16]. It was believed that the risk of pre-term birth, which poses a health threat to both the woman[1] and fetus, could be decreased through the implementation of bedrest [17]. An upright posture was thought to exert excess pressure on the cervix, thus, increasing the likelihood of pre-term delivery [18]. This belief served as justification for the near-unanimous prescription of bedrest in the years to come during pregnancies with multiple fetuses, where cervical pressure is even further amplified. Interestingly, bedrest was a frequently advised intervention at the first sign of any pregnancy complication despite an overwhelming lack of supporting evidence [19].

The rationale for the implementation of bedrest was also partially justified by observations of pregnancy success rates in relation to socioeconomic status. Lower social class often correlated with poorer pregnancy outcomes; thus, it was proposed that higher levels of PA in lower class females were associated with these adverse outcomes [20]. While higher class females often had the means to assume bedrest in late gestation, lower class pregnant females worked until the end of their pregnancies, rendering PA and upright posture unavoidable [20]. PA redirects blood flow to the working musculature, exposing the placenta to a theoretical state of decreased blood flow and therefore depriving the fetus from nutrients and oxygen[20]. Accordingly, it was widely believed that PA was the culprit of many pregnancy complications, and prenatal bedrest was deemed a worthy intervention. This class-based divide in the practice of and access to bedrest was also seen in the postpartum period, where it was a socially encouraged part of Victorian society. Upper class females would endure what was known as the *lying-in* period, where remaining in bed post-birth was strictly enforced, whose duration could be up to several weeks [21]. In reality, the disparity in complication-free pregnancy rates based on socioeconomic status is a complex matter with many confounding variables, including but not limited to maternal nutrition, environmental stressors, smoking and substance use, and social support[22].

Prescribing medical professionals rarely considered the detrimental impacts of bedrest. Inactivity for extended periods of time has harmful effects on health, implicated with increased muscle atrophy and weight loss, decreased cardiovascular function, disrupted hormones, and negative mental health outcomes (see section 2.2) [19,23]. Based on surveys performed by Maloni *et al.* in 1998[23], physicians prescribing bedrest were largely unaware of its associated side effect. Consequently, females who had endured this largely unhelpful and unfounded intervention were too often left post-treatment without proper recovery, education, or guidance for their newly deconditioned bodies.

---

[1] The inclusive term used by our team is gestational parent (gesP), however, we have chosen to use the term woman/women if it was the wording in the original historical text

Today, we know that there is little to no supporting evidence for bedrest as a successful pregnancy intervention, and often does more harm than good[24]. Though historically it was an acceptable practice, research in the last century has made slow progress to uncover the truth about maternal bedrest, PA, and pregnancy, rendering unsupported recommendations for inactivity in healthy pregnancies a thing of the past.

### 2.2. Early 1900s: Emphasis on Rest and Housework.

Following the bed rest and *lying-in* period, emerging practices of pregnancy "hygiene" began in the early 1900's. While much of the literature and the expected behaviours during this time period were purely opinion-based, medical professionals recommended that women should rest as much as they could, including a minimum 30-minute nap per day in addition to their regular chores [25,26]. The aspect of rest was heightened during the later portions of pregnancy. It was thought that pregnancy fatigued the body, especially the heart, and that strenuous exercise may lead to cardiac insufficiency [27]. Moreover, the expanding uterus was thought to misplace the heart and impair the diaphragm such that exercise could not be sustained [28,29].

Throughout the prevailing belief that the pregnant body cannot maintain exercise, very light activities, such as walking outdoors, were recommended [25,26,30,31]. However, since women were responsible for the household at this time, doctors presumed that daily chores would be acceptable [26]. As such, many of the recommendations at this time included quitting all sports to conserve energy not only to support the growing fetus, but to continue performing chore work and upkeep the household. Only a few doctors recommended more "strenuous" activities, such as gymnastics during pregnancy, in addition to walking outdoors [25]. There were no specific guidelines for the duration or volume of exercise at this time, with some literature suggesting no limits as to walking and gymnastics engagement [25].

Of the exceedingly limited scientific research that included women during the early 1900s, little attention was paid to pregnant women, and much of the pregnancy research was focused on the implications of pregnancy and heart disease [28,29,31]. A study completed by Reid in 1930[28] looking at heart disease during pregnancy compared necropsy files from male and female participants. It divided the female participants based on marital status, not parity status; representing a bias that all married females bear children. After evaluating the mitral orifice and cause of death, this study concluded that pregnancy may cause cardiac insufficiency[28]. Since the participants were not actively pregnant at the time of death, nor were they divided by the number of pregnancies and time after birthing children, many confounding variables could explain the increase in cardiac insufficiency observed in the married group. Additionally, Newell (1912)[31] remarked that when pregnant women who did not engage in exercise came into labour, acute left ventricular dilation was often observed. Therefore, Reid's findings of cardiac insufficiency may not have been the result of pregnancy, but potentially due to the lack of engagement in exercise and PA throughout pregnancy. In support of exercise, other doctors observed a decrease in the prevalence of toxemias (preeclampsia as we know it today) and other diseases when women were considered "fit" [25,31,32]. Interestingly, it was suggested that exercise and careful dieting during pregnancy may reduce the risk for obstetrical and surgical interventions during labour [30].

In summary, research from the early 1900s supported engagement in light exercise and chores; however, prioritizing rest throughout pregnancy remained the standard practice. The expectations for pregnant women at this time were heavily routed in social norms and gender roles to support the household. Accordingly, researchers may have been influenced by these cultural standards and the worry of harming the fetus stymieing further exploration of the benefits of exercise during pregnancy.

### 2.3. *1950s - late 1970s: Training for Labour Era*

In the 1950s, the publication of Helen Headman's book, *A Way to Natural Childbirth,* represents the first time exercise classes focused on labour training appeared in the literature. The popularity of this concept led doctors to recommend exercises and classes that physically prepared and educated pregnant women for pregnancy and labour[33]. The physical exercise portion of this training for labour program was two-fold: (1) to increase the tone and efficiency of the core muscles and the pelvic floor and (2) to promote relaxation. Similar training programs emerged during this era to help prepare for labour.

As the view of women being responsible for the house was still commonly held in the mid-1900s, exercise engagement in the form of walking outdoors was still recommended to ensure women could continue to fulfill their housekeeping duties during pregnancy, despite some of the literature noting that housework was no longer a substitute for exercise[34]. These training programs were not simply aimed to assist in labour; but rather it was thought that if abdominal strengthening began during the second trimester and into the post-partum period, women would more rapidly regain their figure and experience a reduction in the risk of uterine prolapse[34]. The recognition of the importance of the pelvic floor, and what are known today as Kegel exercises, emerged through the work of Arnold Kegel. He described a set of pelvic floor exercises that can be done with a *Perineometer* to restore function and tone to the pelvic floor during the post-partum period[35]. The maintenance of proper posture during pregnancy was also highlighted to reduce the pressure on the pelvis, support proper circulation, and reduce lower back pain[36]; but this literature was not based on research evidence, just medical opinion.

In addition to strengthening the core and pelvic floor, exercise classes termed "ante-natal" classes or "Lamaze" classes gained popularity and were the outcome of the exercise boom in the 1970's [37,38]. These classes still maintained the common theme for this era of preparing for birth, but added the aspect of breathing techniques and strengthening the entire body. In a hallmark study conducted by Gunter (1956) [38] that looked at the use of ante-natal classes and the implications for birth, the exercise group laboured on average 6 hours less, experienced fewer episiotomies, half the number of birth complications and less post-partum hemorrhage compared to the control group [38]. Later in the 1960s and 70s, additional studies supported the benefits of exercise, demonstrating that fit women had shorter labour times and fewer complications than unfit women[39,40].

Outside of the training for labour framework that was commonly denoted in the literature during the 70s, researchers started to examine the physiological responses of exercise in the pregnant population [41,42]. Pregnant women were shown to have increased ventilation, heart rate, and cardiac output per unit increase in work as compared to non-pregnant controls [41]; thus illustrating that pregnant women can sustain exercise. Moreover, data emerged linking exercise during pregnancy to better management of pre-pregnancy diabetes [43].

Popular medical opinion from the late 19$^{th}$ and into the first decades of the 20$^{th}$ century was that pregnant women should use extreme caution to avoid fatigue and overexertion. The majority of the early published guidelines for gesPs were unscientific and reinforced the notion that females were weak and frail. As true scientific research began to emerge in the 70s, the ability to sustain exercise during pregnancy, along with the potential benefits, started to come into view.

### 2.4. *1980s – Present Day: Evaluating the Risks and Benefits*

While cardiopulmonary responses to exercise in pregnant females had been reported in the 1970s, the 1980s represented a boom in exercise physiology research in pregnancy. As much remained unknown, there were persistent concerns surrounding exercise risks in pregnancy; including the threat of a decrease in oxygen and nutrient delivery to the fetus, a fear of hyperthermia, and increased stress on the fetus initiating preterm labour and potential fetal mortality [20,44,45]. Albeit, the risks were a cause for concern at the time, the literature commonly espoused that the benefits of exercise outweighed the risks [44]. Thus, exercise with moderation (heart rate monitoring) was promoted with the caveat that exercise was not appropriate for those carrying more than one fetus or with a predisposition to heart disease[44].

One study that contributed immensely to the concerns of fetal mortality with exercise was by Briend (1980)[20], whose group examined pregnant females who worked physical jobs that required a lot of standing. They found that maternal mortality and stillbirth rates were higher in working pregnant women compared to those who did not work [20]. It is important to note that this study can be criticized for sampling bias and confounding variables as previously discussed in section 2.1.1.

Many historical myths surrounding the dangers of exercise in pregnancy have been disproven with evidence showing no increase in neonatal mortality and obstetrical complications, no detrimental changes in oxygen and nutrient availability and thus, no negative implications on fetal growth and development with gestational exercise [46,47]. There was also very little evidence that exercise led to changes in the fetal metabolism or blood catecholamines, signifying no changes to fetal stress with exercise [47]. Concerns of hyperthermia, fetal hypoxia and deprivation of nutrients and pre-term labour were also valid due to the lack of knowledge regarding exercise in pregnancy. Some researchers viewed these gaps in the literature as an opportunity to dive into the field of exercise physiology focused on pregnant females. Among these scientists was James F. Clapp III, who undertook a suite of hallmark studies that advanced exercise in pregnancy research. One integral study in the early 1990s performed by Clapp (1991)[48] provided evidence that core temperature increases during exercise were not detrimental to the embryo and fetus due to the physiological adaptations during pregnancy. Other studies also supported Clapp's findings that the changes in core temperature experienced during exercise were very low and did not jeopardize the fetus [47].

There was a common trend that emerged with some of Clapp's research showing decreased weight gain during pregnancy and lighter weight offspring compared to non-exercising individuals [49–51]. Specifically, a study by Clapp and Capeless (1990) [49] found significant reductions in birth weight, fetal weight and size percentiles along with decreased adiposity without differences in crown-to-heel length and head circumference in the exercise group compared to non-exercising controls. The main differences (~70%) in weight reduction were thought to be equated to the reduced adiposity of the offspring [49]. While these changes in birth weight did reach statistical significance, this study has been criticized in that the reduction in weight was within normal ranges with no concern for preterm birth. Contrarily, when pregnant women who did not previously exercise were randomized to a weight-bearing exercise group for 8 weeks, their offspring were found to be heavier and longer than non-exercising controls[52]. This difference in weight was thought to be caused by an increase in both the offspring's lean mass and fat mass [52]. Interestingly, this study also found a greater placental growth rate and indexes of placental function during mid-pregnancy in the exercise group [52]. With these results in mind, it was concluded that aerobic exercise was safe for pregnant females and the fetus.

Additional benefits of exercise during pregnancy unearthed in the early 90s included reduced blood pressure, decreased risk of CVD, and the management of pre-pregnancy diabetes[44]. With the emergence of a greater volume of research with the transition to the 21st century, it became more apparent that exercise during pregnancy posed no threat for low-risk pregnancies, showed no detrimental effects and, in many cases, was beneficial for the pregnant woman and fetus. There was a continuation from the other decades in the observations of lower incidence of vaginal and abdominal surgeries along with shorter labour times with pregnant women who exercised[49]. Adding on to these observations, less fetal stress was observed in women who exercised compared to those who did not exercise throughout pregnancy[49].

Later in the 1980s, the publication of the Fetal Origins Hypothesis (FOH) paradigm by Sir David Barker, known today as the Developmental Origins of Health and Disease (DOHaD) paradigm, opened the door to examine PA and exercise as an *in-utero* exposure that may impact the health of the fetus. While Dr. Barker interests did not lie with exercise, his seminal work examining geographical and temporal patterns of disease, unveiled a relationship between heart disease and neonatal and post-neonatal mortality in the same locations 70 years prior[53]. He asserted that both intrauterine and early life factors influenced the risk of disease development in later life[53] Much of the current work on the DOHaD paradigm has focused on factors such as nutrition, exposure to chemicals, smoking, and disease status of the pregnant female[54] (Figure 2); there has been limited research on the prevention of these adverse outcomes through environmental exposure like prenatal PA and exercise.

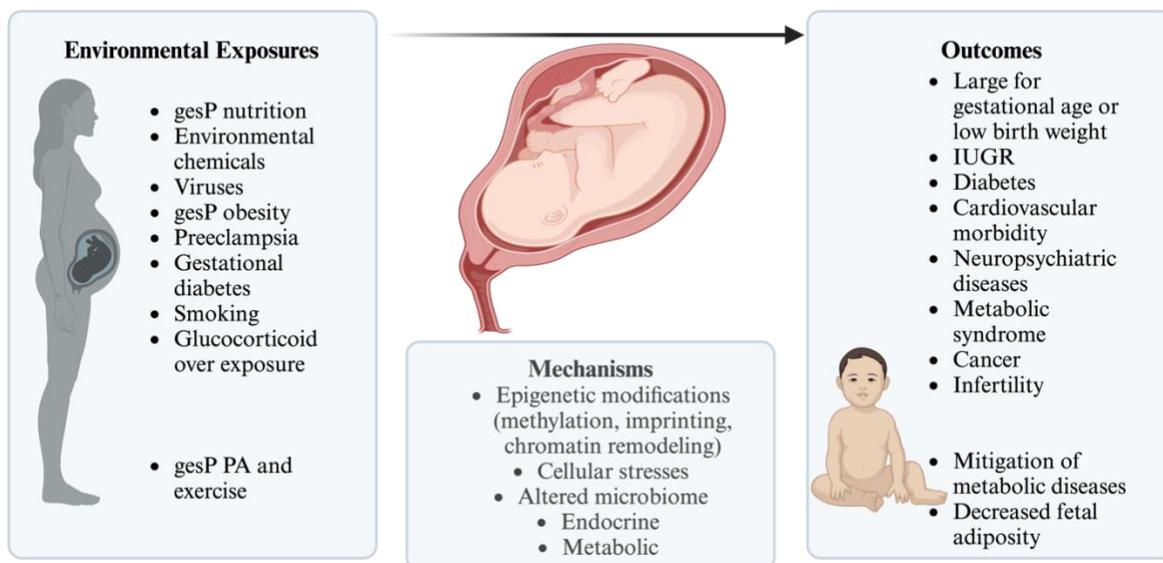

Figure 2. Current findings supporting the DOHaD paradigm. gesP; gestational parent: PA; PA: IUGR; Intrauterine growth restriction. Created using BioRender.com Sources: [54–62]

Guidelines for exercise during pregnancy also began to be featured in the literature from the American College of Obstetrics and Gynecologists (ACOG) starting in 1985. The ACOG promoted that most aerobic exercise is safe for pregnancy and that pre-conception exercise can be maintained during pregnancy and the post-partum period [63,64]. These largely opinion-based recommendations were carried over to the early 2000s [65] when it was thought that exercise had not yet conclusively shown to be beneficial for improving perinatal outcomes [63]. Believing previous guidelines to be too conservative, and a desire to highlight the abundance of research over the last two decades showing no increases of early pregnancy loss, pregnancy complications, abnormal fetal growth, or negative fetal outcomes [66] a Canadian team, undertook the task of creating the first set of evidence-based clinical guidelines for PA in pregnancy. With the collaboration of the Society of Obstetrics and Gynaecologists Canada and the Canadian Society of Exercise Physiology (CSEP), the *2019 Canadian Guideline for Physical Activity Throughout Pregnancy* were developed using a vigorous methodological approach[67]. Recommendations put forth in the updated guidelines were supported by a set of 12 systematic reviews aimed at identifying the characteristics of exercise (i.e., frequency, intensity, duration, type, and volume) that were favourably associated with maternal, fetal, and neonatal health outcomes (summarized in Figure 3)[67]. The recommendations are that all women without contraindications, including those with GDM, were previously inactive, or are classified as a person living with obesity or overweight, should engage in a minimum of 150 minutes of moderate-intensity PA each week over a minimum of 3 days per week, although activity is encouraged every day.

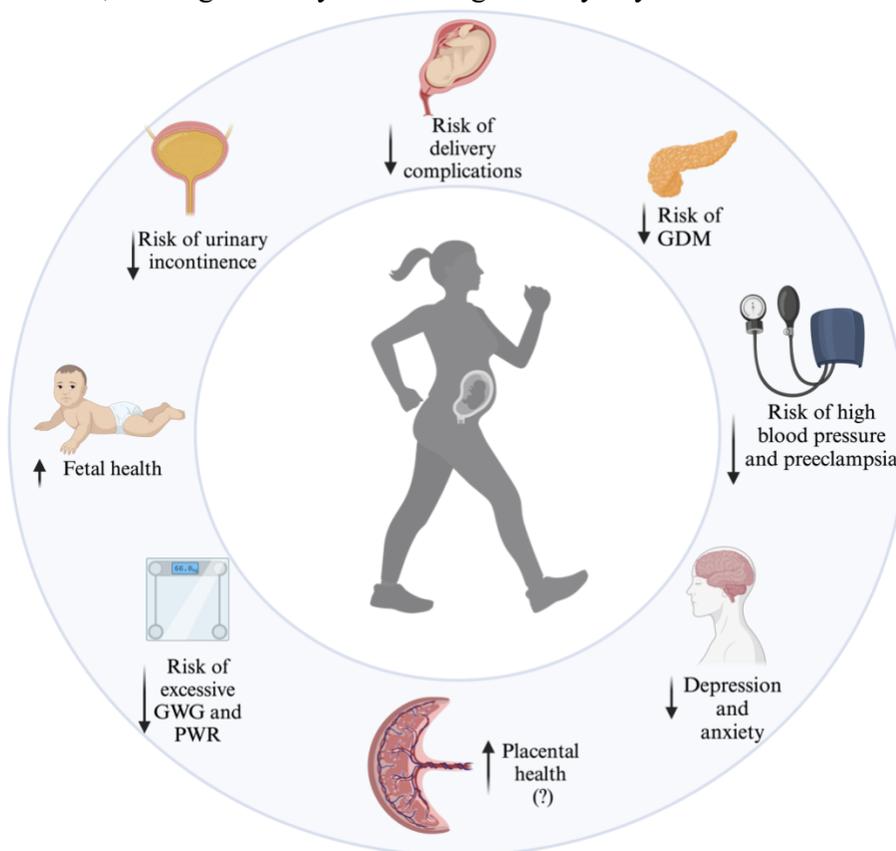

Figure 3. Benefits of Gestational Parent PA. Adapted from Bhattacharjee et al. (2021)[68]. GDM; gestational diabetes mellitus: GWG; Gestational weight gain: PWR; postpartum weight retention. Created with BioRender.com

There were vast advances in research on pregnancy and exercise during the 1980s to early 2000s to evaluate the risks and benefits of exercise during pregnancy along with hallmark studies that are the basis for current-day studies and advances in clinical and research obstetrics (Figure

4). While the field of exercise and pregnancy has grown immensely since the 1850's leading to science-informed exercise during pregnancy guidelines[67], there are still gaps in knowledge that need to be filled. While researchers have observed the fruitful benefits of PA during pregnancy, the mechanisms through which these benefits are accrued are not yet fully elucidated. Some of these mechanisms may include changes to oxygen tension in the placenta, remodelling of structures at the gesP-fetal interface, or changes in nutrient transport, cytokines and myokines. For a more in-depth review of the physiological responses to PA in pregnancy, please see other research work by the Adamo team[68–70].

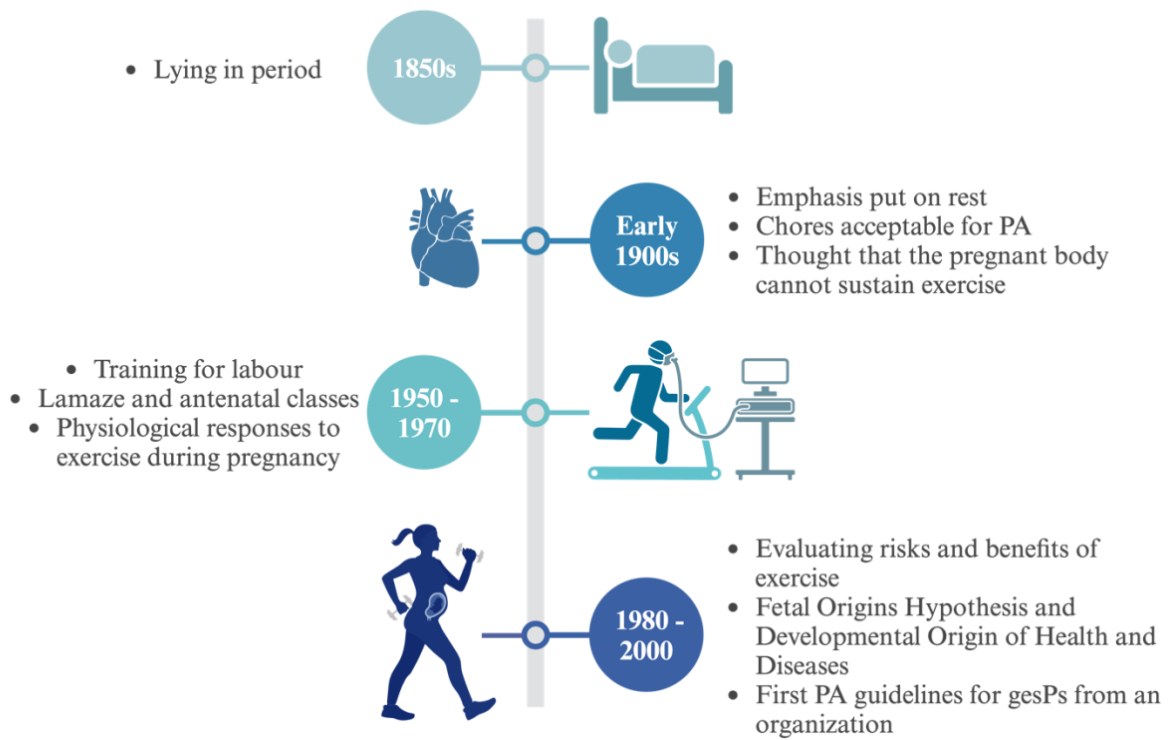

Figure 4. The Evolution of Exercise Physiology Research in Pregnant Populations. PA: PA; gesP: gestational parent. Created with BioRender.com

## 3. Future Directions
### 3.1 *Mental Health and Pregnancy*

It is increasingly clear that the mental health of the gesP plays an integral role in long-term outcomes for both the birthing parent and the fetus. Prenatal depression is a strong predictor of postpartum depression [71], a condition impacting approximately 17% of gesPs globally [72]. As such, gestational depression, anxiety and stress can have lasting effects on fetal development. Offspring from birthing parents with high levels of prenatal stress have greater risk of decreased cognition, emotional problems such as depression and anxiety, sleep problems, as well as physical and physiological impairments (the reader is referred to Glover (2014)[73] for a more encompassing review). One study comparing the lipidome of placentas of gesPs with and without prenatal depression found several long-chain polyunsaturated fatty acids (LC-PUFA), including docosahexaenoic acid (DHA), to be down-regulated in depression[74]. DHA is critical for fetal brain development, and reduced levels in offspring at birth have been associated with decreased cognition at seven years of age [75]. It has also been theorized that the high cortisol levels that are often associated with poor mental health contribute to its negative impacts on offspring [73,76]. While the physiological alterations resulting from poor mental health during pregnancy are not yet understood, it is evident that gesP mental health fosters a mutually harmful environment for them and the fetus.

Despite their frequent prescription in the past, bedrest and sedentary behaviour have been associated with a plethora of negative mental health implications in pregnancy. Studies ranging from prolonged bedrest in postoperative patients to sedentary university students consistently demonstrate higher rates of anxiety, stress and depression [19,77,78]. Specific to pregnancy, gesPs have been shown to experience several psychological postpartum symptoms following antepartum bedrest, with the length of rest significantly correlated to the number of postpartum symptoms expressed [79]. Bedrest is further implicated with high depressive symptoms and antepartum stressors during multifetal pregnancies, continuing into the first several weeks of the postpartum period [80]. While antenatal mental health should be of critical importance, it has not frequently been highlighted or appropriately addressed in the literature. As a fitting reflection of the historically prominent bias toward men as study participants in scientific literature, some of the first papers to address the mental health implications of bedrest in pregnancy focused on the male partner, and the stressful environment they endured while the gesP could not contribute to the household [81,82].

Recently, PA has been shown to positively effect mental health in the general population and as research progresses, similar patterns are emerging for gesPs. In 2019, a systematic review found significantly reduced postpartum depression scores in physically active gesP. The effect sizes varied greatly across the included studies, suggesting confounding effects of frequency, intensity, type and duration of exercise [83]. Therefore, exercise guidelines should be explored to optimize mental health during pregnancy. Similarly, a 2022 systematic review of observational studies pertaining to PA and gesP mental health found PA to be associated with lower rates of prenatal depression, anxiety and stress, and decreased risk of postnatal depression and anxiety[84]. Interestingly, the same review noted no association of PA prior to pregnancy on these outcomes, though only a small number of studies exist on this topic.

While multiple umbrella and systematic reviews support PA as having a favourable consequence on gesP mental health, many agree that additional high quality evidence is required to elucidate this relationship fully [83–86]. As science progresses, we have transitioned to a stage where bedrest, formerly a standard pregnancy recommendation, is now understood to be harmful to mental health. The literature is uncovering factors that effect gesP mental health and exploring PA as a viable intervention. Future work should continue to analyze the repercussions of PA in this context and examine the role of potential moderators such as PA variables (frequency, intensity, time and type of activity) and pre-pregnancy activity levels.

### 3.2. Specialty populations

While international guidelines endorse engaging in at least 150 min/week of moderate-intensity PA across pregnancy for uncomplicated pregnancies, we have limited knowledge about the upper and lower boundaries of exercise frequency, intensity, type and time (duration). Pregnant individuals can inhabit extreme categories: high-performance expectations (i.e., elite athletes) and extremely low performance (i.e., hospitalization for complications), and there are research gaps to fill on both ends of this activity spectrum.

#### 3.2.1 Athletes & Arduous Occupations

Given that females constitute half of the global population, there is a pressing need to include, and encourage female participation in sports and exercise science research[87]. As the number of females participating in a wide variety of training modes and modalities (e.g., HIIT, Pilates, CrossFit, weightlifting) continues to increase, adequate research funding must be allocated to ensure equity in understanding the unique aspects of female physiology and performance across exercise disciplines. Female athlete representation at elite levels has grown exponentially, exemplified by their record-breaking presence at the recent Tokyo Olympics[88]. Notably, the peak performance years for female athletes often align with their prime reproductive years, making it inevitable that pregnancies will intersect with training and competitive periods. Despite this knowledge, there remains a dearth of research and frameworks specifically tailored to address the needs of female athletes, particularly in the antenatal period[89]. As there is limited evidence available regarding pregnant athletes who engage in high-intensity exercise, current clinical recommendations[90,91] as well as the consensus statement of the International Olympic Committee[92] rely heavily on expert opinions and best practices.

In parallel, the representation of females in arduous occupations (e.g., military, law enforcement, firefighting, manual labour) is increasing and, like athletes, many of these individuals are in their reproductive years. Occupational exposures (e.g., chemicals, stress, physical assault, shift work, biological agents) combined with the physiological, anatomical, and biomechanical changes associated with female reproduction result in numerous health and safety concerns for pregnant, post-partum, and parous females employed in arduous occupations [93]. Females returning to military service postpartum are known to be at an increased risk of musculoskeletal injuries [94], and female servicemembers who have given birth sustain more repetitive strain injuries than nulliparous peers[95]. Yet, when provided support to overcome the potential challenges associated with childbearing, pre-pregnancy physical performance can be regained and surpassed. For example, elite runners who continue training over pregnancy are able to return to or exceed pre-pregnancy performance 1-3 years post pregnancy.[96] Similarly, if military members are able to overcome the fitness deficits and increased health risk seen in the first 12-18 months postpartum[97], they are capable of attaining equal or better physical fitness than nulliparous matched peers. Significant knowledge gaps persist due to the insufficient research on pregnancy, postpartum recovery, and resumption of high physical exertion among both elite athletes and those employed in arduous occupations (e.g., military service members, police officers, firefighters, manual labourers)[90]. While exercise guidelines include specific recommendations for the general population[67] and recreational athletes[98], guidance for pregnant athletes engaging in high volume or high-intensity training remains scarce. There are, however, studies indicating that women can sustain high performance levels during pregnancy[99] and regain their pre-pregnancy fitness following childbirth[96], but further research with larger sample sizes examining the limitations and required obstetrical management in these populations will enable a safer and more prompt return to work or competition.

### 3.2.2. High risk pregnancies - is activity restriction and bedrest the only option?

Steeped in tradition, activity restriction and in severe cases, bedrest or hospitalization are common approaches for managing contraindications such as preterm premature rupture of membranes, vaginal bleeding with or without placenta previa, multiple gestation, hypertensive disorders of pregnancy, short cervical length, and fetal growth restriction. This practice flies in the face of increasing evidence, reviewed by Palacio and Mottola[100], indicating that restricting activity does not necessarily prevent negative perinatal outcomes, and might worsen both physical and psychosocial risks.

Knowing that a physically active pregnancy (accruing 150 min/week of moderate-intensity PA) offers health benefits to both the gesP and their offspring and that evidence indicates that excessive sedentary behaviour during pregnancy increases chronic disease risk and may impair birth outcomes[101–103], how do we treat those who experience what are deemed absolute contraindications? In brief, few interventions have been tested to reduce the deconditioning impacts of activity restriction in pregnancy[24] and thus we are presently unsure. What we do know is that severe activity restriction is not the answer. Future research should evaluate the use of low intensity activity routines to lessen the effects of activity-restriction, ensuring gesP are prepared to undertake newborn care responsibilities when returning home following childbirth.

## 4. Conclusion

For many years females have been neglected from medical and exercise physiology research, a trend exacerbated when considering pregnancy. Popular medical opinions from the late 19$^{th}$ century into the first decades into the 20$^{th}$ century were that gesPs should use extreme caution when engaging in PA and exercise to avoid fatigue and overexertion. The bulk of early published guidelines were unscientific, thus hardening the belief that females were weak and frail. These guidelines remained an unquestioned dogma for decades while reinforcing gender norms. Since the 2000s, exercise and PA have consistently shown positive associations with the health of pregnant parents and the fetus. However, large knowledge gaps remain. The deliberate exclusion of pregnant and postpartum females and athletes, along with the inadequate representation of females overall, is unacceptable. With record levels of females entering arduous occupations and participating in elite sporting events, it is imperative that concerted efforts are undertaken across various levels to ensure that research participants accurately reflect the demographics of the population to which the findings are intended to apply.

**Funding:** Both MM and AC are recipients of Ontario Graduate Scholarship (OGS) awards as well as the Canada Graduate Scholarship Awards from CIHR.

**Conflicts of interest:** The authors have no conflicts of interest to declare.

**Figure Captions**

Figure 5. Visualization of the Approximate Volume of Knowledge in Exercise Physiology Research of the Biological Sexes and Gestational Parents. Created using BioRender.com

Figure 6. Current findings supporting the DOHaD paradigm. gesP; gestational parent: PA; PA: IUGR; Intrauterine growth restriction. Created using BioRender.com. Sources: [54–62]

Figure 7. Benefits of Gestational Parent PA. Adapted from Bhattacharjee et al. (2021)[67]. GDM; gestational diabetes mellitus: GWG; Gestational weight gain: PWR; postpartum weight retention. Created with BioRender.com

Figure 8. The Evolution of Exercise Physiology Research in Pregnant Populations. PA: PA; gesP: gestational parent. Created with BioRender.com